\documentclass{iopart}
\usepackage{graphicx}
\usepackage{amsfonts}
\usepackage{amssymb}
\usepackage{mathrsfs}
\usepackage{amsthm}
\usepackage{mathrsfs}
\usepackage{eufrak}
\usepackage[mathscr]{eucal}

\begin{document}

\newtheorem{twr}{Theorem}
\newtheorem{lem}[twr]{Lemma}
\newtheorem{df}[twr]{Definition}
\newtheorem{ex}[twr]{Example}
\newtheorem{pr}[twr]{Proposition}
\newtheorem{fct}[twr]{Fact}

\title{General theory of detection and optimality}
\author{G Sarbicki}
\address{Insitute of Physics, Nicolaus Copernicus University \\
Grudziadzka 5, 87--100 Toru\'n, Poland}
\ead{gniewko@fizyka.umk.pl}

\begin{abstract}
 A general formulation of the problem of detection for a pair of two cones is presented. The special case is the detection of entangled states by entanglement witnesses. Having defined what means ``to detect'', one can identify the subset of elements, which detect optimally. I will present the properties of this set for a general pair of cones.

 In particular, I prove the generalization of the theorem of Lewenstein, Krauss, Cirac, Horodecki. The entanglement witness $W$ is optimall if the set of product vectors $\{ \phi \otimes \psi: \langle \phi \otimes \psi | W | \phi \otimes \psi \rangle = 0\}$ spans the whole Hilbert space of a system. There exist optimall entangled witness, which do not fullfill this property. It is closely related to some geometrical properties of the boundary of the set of entanglement witnesses and it is possible to say something more about location of such extraordinary states.
\end{abstract}

\section{Introduction}

In the set of mixed states of a bipartite quantum system one can define the subset of separable states \cite{Wern}. The state $\rho$ is called separable, when there exists a decomposition $\rho=\sum_i p_i \rho_i^{(1)} \otimes \rho_i^{(2)}$. In such states of a system, its subsystems can be correlated only clasically.

There is no general method to determine, whether a given state is separable or not. One of the most important tools are entanglement witnesses \cite{sep3H}, \cite{Terh}. A hermitian observable is called entanglement witness, when its mean value in any separable state is positive, but the observable is not semi-positive.

For an entanglement witness $W$ we can define the set of entangled states, which are \textit{detected} by this entanglement witness, i.e. the states in which the mean value of the entanglement witness is negative. We denote this set by $\mathcal{D}(W)$. Now, we say that entanglement witness $W_1$ is \textit{finer} than an entanglement witness $W_2$, when it detects more states, i.e. when $\mathcal{D}(W_1) \supset \mathcal{D}(W_2)$. Entanglement witness, for which there exists no finer witness, is called \textit{optimal} \cite{opt}.

In the set of quantum states we can define another set of states, the set of PPT states. A states is a PPT state, when its partial transposition $\rho^\Gamma = (I \otimes T) \rho$ is positive. The set of PPT states is a superset of the set of separable states, and the equality holds only, when the dimensions of subsystems are $2 \times 2$ and $2 \times 3$ \cite{sep3H}. All entangled PPT states are bound entangled \cite{PPTbound}. Entanglement witness, which cannot detect PPT entangled states, is called \textit{decomposable} \cite{opt}. Any decomposable entanglement witness can be written as $W=A+B^\Gamma$, where $A$ and $B$ are semi-positive. Entanglement witnesses, which are not decomposable, are called \textit{non-decomposable}.

A nondecomposable entanglement witness $W_1$ is called \textit{nd-finer} (non-decomposable finer) than another non-decomposable entanglement witness $W_2$, when $W_1$ detects more PPT enatangled states than $W_2$. A non-decomposable entanglement witness $W$ is called \textit{nd-optimal} (non-decomposable optimal), when there is no other entanglement witness detecting more PPT entangled states than $W$ \cite{opt}.

There are two theorems characterizing optimality \cite{opt}:

\begin{twr}
 $W_1$ is finer than $W_2$ $\iff$ $W_2 = \lambda W_1 + P$, for a positive scalar $\lambda$ and a semi-positive observable $P$.
\end{twr}

\begin{twr}
 $W_1$ is nd-finer than $W_2$ $\iff$ $W_2 = \lambda W_1 + D$, for a positive scalar $\lambda$ and $D = A + B^\Gamma$ for semi-positive observables $A$ and $B$.
\end{twr}

I will present later a generalization of these two theorems. To do that, it is necessary to remind some basic concepts and facts about the geometry of proper cones.

\section{Geometry of proper cones}

This section presents basic definitions and facts of theory of proper cones. For a more detailed discussion see \cite{Alipr}, \cite{Barv}.

\begin{df}
 A set $K \subset \mathbb{R}^N$ is called a proper cone, iff:
\begin{enumerate}
 \item $\forall \mu, \nu \ge 0 \quad \forall x,y \in K \quad \mu x + \nu y \in K$
 \item $K$ is closed in $\mathbb{R}^N$
 \item $\mathrm{span}K = \mathbb{R}^N$ (fullness)
 \item There exists no subspace of $\mathbb{R}^N$ contained in $K$ (pointedness)
\end{enumerate}
\label{propcone}
\end{df}
A set of points of a cone differing by a positive scalar is called a \textit{ray} of the cone. Any ray can be written as $\{ k \cdot x: \ k \in \mathbb{R}_+ \}$, and then we say, that it is generated by an element $x$. A ray is called \textit{extremal}, when a point of the ray cannot be decomposed as a convex combination of points out of the ray.

\begin{ex}[Examples of proper cones]

The set of unnormalized quantum states (positive matrices) of $d$-level system is a proper cone in the real vector space of hermitian matrices $\mathcal{B}(\mathbb{C}^d)$. It's extreme rays are generated by projectors of rank one. This cone will be denoted by $\mathcal{B}_+(\mathbb{C}^d)$ or simply by $\mathcal{B}_+$.

A matrix $\rho \in \mathcal{B}_+(\mathbb{C}^{d_1} \otimes \mathbb{C}^{d_2})$ is called unnormalized separable state of two subsystems of dimensions $d_1$ and $d_2$, when it can be decomposed as
\begin{displaymath}
\rho = \sum_i A_i \otimes B_i,
\end{displaymath}
where $A_i \in \mathcal{B}_+(\mathbb{C}^{d_1})$ and $B_i \in \mathcal{B}_+(\mathbb{C}^{d_2})$. It's easy to check, that the set of unnormalized separable quantum states is a proper cone in the space of hermitian matrices $\mathcal{B}(\mathbb{C}^{d_1} \otimes \mathbb{C}^{d_2})$. An extreme ray of this cone is generated by a tensor product of rank-one semi-positive matrices, so one can write an alternative definition of unnormalize separable state as:
\begin{equation}
 \rho = \sum_i |\phi_i \otimes \psi_i \rangle \langle \phi_i \otimes \psi_i|,
 \label{sep-st}
\end{equation}
where vectors $\psi_i$ and $\phi_i$ need not to be normalized.
This cone will be denoted by $\mathcal{S}_1(\mathbb{C}^{d_1} \otimes \mathbb{C}^{d_2})$ or simply by $\mathcal{S}_1$.

A bipartite quantum state $\rho$ is called PPT state, when $\rho^\Gamma \ge 0$. The set of unnormalized PPT states is an intersection of cones $\mathcal{B}_+(\mathbb{C}^{d_1} \otimes \mathbb{C}^{d_2})$ and $\mathcal{B}_+^\Gamma (\mathbb{C}^{d_1} \otimes \mathbb{C}^{d_2})$. The set of its extreme rays is not known in general. This cone will be denoted by $\mathcal{S}_{PPT}(\mathbb{C}^{d_1} \otimes \mathbb{C}^{d_2})$ or simply by $\mathcal{S}_{PPT}$.

 The set of positive matrices and witnesses detecting entanglement in $d_1 \times d_2$-level quantum system (a set of matrices positive on product vectors) is a proper cone in $\mathcal{B}(\mathbb{C}^{d_1} \otimes \mathbb{C}^{d_2})$. The set of its extreme rays is not known in general. This cone will be denoted by $\mathcal{W}_1(\mathbb{C}^{d_1} \otimes \mathbb{C}^{d_2})$ or simply by $\mathcal{W}_1$.

 The set of positive matrices and decomposable witnesses in $d_1 \times d_2$-level quantum system is a proper cone, and its extreme rays are generated by matrices of the form $P$ or $P^\Gamma$, where $P$ is a projector of rank one. This cone will be denoted by $\mathcal{W}_D(\mathbb{C}^{d_1} \otimes \mathbb{C}^{d_2})$ or simply by $\mathcal{W}_D$.
\end{ex}

\subsection{Duality}

For a cone $K$ in a real vector space $X$ one defines a proper cone $K^*$ in $X^*$.
\begin{df}
 A set $K^*$ defined as
\begin{displaymath}
 K^* = \{ y \in X^*: \quad \forall x \in K y(x) > 0 \}
\end{displaymath}
is called a dual cone of a proper cone $K$.
\end{df}
One can restrict the quantified set in definition to points of extreme rays of $K$.

The set $K^*$ is a proper cone. One can consider then the proper cone $(K^{*})^{*}$. Using the reflexivity of a finite-dimensional real vector space, one can easily prove, that $(K^{*})^{*} = K$. The duality of proper cones has the following properties:
\begin{fct} The properties of duality of cones:
 \begin{itemize}
  \item $K \subset L \ \Rightarrow \ K^* \supset L^*$.
  \item $(K \cap L)^* = \mathrm{conv}(K^* \cup L^*)$.
  \item $\mathrm{conv}(K \cup L)^* = K^* \cap L^*$
 \end{itemize}
\label{dualprop}
\end{fct}

An inner product in $X$ constitutes isomorfism between $X$ and $X^*$. One can then consider $K$ and $K^*$ as elements of the same space. When for a cone $K$ holds $K=K^*$, one calls $K$ \textit{self dual}. In spaces of hermitian matrices, which we are interested in, such an inner product is Hilbert-Schmidt product.

\begin{ex}[Quantum states]
 The cone $\mathcal{B}_+$ is self dual. Indeed, a matrix $\rho \in \mathcal{B}_+$ is semi-positive, iff $\forall \psi \ \langle \psi | \rho | \psi \rangle \ge 0$, what can be rewritten as $\forall \psi \ \mathrm{Tr}(| \psi \rangle \langle \psi | \rho) \ge 0$. The matrix $\rho$ is then positive on all extreme rays of $\mathcal{B}_+$, so $\rho \in \mathcal{B}^*_+$.
\end{ex}

\begin{ex}[Separable states and entanglement witnesses]

By definition, the matrix $W$ is an element of the proper cone $\mathcal{W}_1$ iff $\forall \rho \in \mathcal{S}_1 \ \langle \rho | W \rangle_{HS} \ge 0$, so $W \in \mathcal{S}^*_1$. These proper cones are dual to each other.
\end{ex}

\begin{ex}[PPT states and nd-witnesses]
 The proper cone $\mathcal{S}_{PPT}$ is an intersection of two proper cones: $\mathcal{B}_{+}$ and $\mathcal{B}^\Gamma_{+}$. Due to the second property in Fact \ref{dualprop}, the proper cone $\mathcal{B}^*_{PPT}$ is a convex hull of the sum of proper cones: $(\mathcal{B}_{+})^* \ \cup \ (\mathcal{B}^\Gamma_{+})^* = \mathcal{B}_{+} \ \cup \ \mathcal{B}^\Gamma_{+}$. Such a sum is spanned by the sum of sets of extreme rays of both cones, so its extreme rays are generated by matrices $P$ and $P^\Gamma$, where $P$ is a projector of rank one. The proper cone spanned by the set of such extreme points is $\mathcal{W}_{D}$. The cones $\mathcal{S}_{PPT}$ and $\mathcal{W}_{D}$ are dual to each other.
\end{ex}

\subsection{Faces of a cone}

A subset $F$ of a proper cone $K \subset \mathbb{R}^N$ is called a \textit{face} of a cone, if it is an intersection of the cone and a kernel of a linear functional which is non-negative on the cone. Geometrically, a face is an intersection of the cone and a hipersurface tangent to the cone. The fact that $F$ is a face of a cone $K$ is denoted by $F \vartriangleleft K$.

A face $F$ of a proper cone $K$ is a proper cone in subspace $\mathrm{span} F$. A face $G$ of this cone is also a face of $K$. It allows us to define a relation in the set of faces of the cone $K$:

\begin{df}[Subface]
 Having given two faces $F,G \vartriangleleft K$, we call a face $G$ a \textit{subface} of a face $F$, iff the face $G$ is a face of the proper cone $F$ in the subspace $\mathrm{span} F$.
\end{df}

The relation of beeing subface constitutes a partial order in the set of faces of a proper cone. The maximal element due to this partial order is the whole cone $K$, and the minimal element is the face $\{ 0 \}$.

Intersection of two faces is a face. It allows us to define for a given $x$ the minimal face containing $x$ as the intersection of all faces containing $\rho$. Such a face is said to be generated by an element $\rho$. From now, the face of a cone $K$ generated by an element $x$ will be denoted by $F_K(x)$.


A subset $G$ of a proper cone $K$ can be mapped to a subset $\mathfrak{C}_K(G)$ of the dual cone $K^*$ via formula:
\begin{displaymath}
 \mathfrak{C}_K(G) = \{ y \in K: \ \forall g \in G \ y(g) = 0 \}
\end{displaymath}
It is easy to check, that $\mathfrak{C}_K(G)$ is a face of the cone $K^*$. What's more $\mathfrak{C}_K(G) = \mathfrak{C}_K(F_K(G))$. The mapping $\mathfrak{C}_K$ maps faces of $K$ to faces of $K^*$. Face $\mathfrak{C}_K(F)$ is said to be \textit{complementar} to face $F$:

\begin{df}[Complementary face]
 A face $\mathfrak{C}_K(F) \vartriangleleft K^*$ defined by the formula
\begin{displaymath}
 \mathfrak{C}_K(F) = \{ y \in K^*: \ \forall x \in F \ y(x) = 0 \}
\end{displaymath}
is called a \textit{complementary face} of the face $F$.
\end{df}

Further, we will need some properties of complementarity:

\begin{pr} Properties of complementarity:
\begin{enumerate}
 \item $F \vartriangleleft G \ \Leftrightarrow \ \mathfrak{C}_K(G) \vartriangleright \mathfrak{C}_K(F)$.
 \item $\mathfrak{C}_K(\{ 0 \}) = K$.
 \item $\mathfrak{C}_K(K) = \{ 0 \}$.
\end{enumerate}
\end{pr}

\begin{ex}[Faces of a cone of positive matrices]
 The structure of the cone $\mathcal{B}_+(\mathbb{C}^d)$ is exactly known \cite{faceB+}. A face generated by a given element $\rho \in \mathcal{B}_+(\mathbb{C}^d)$ is the set of all positive matrices with the image contained in the image of $\rho$. The dimension of $F_{\mathcal{B}_+}(\rho)$ is equal $(\mathrm{rank}\rho)^2$. Faces are then in one-to-one correspondence with the lattice of subspaces of $\mathbb{C}^d$. Denote the face of matrices with the image contained in a subspace $V$ as $F_V$. It's quite easy to find the face complementary to $F_V$. We have $\mathfrak{C}_{\mathcal{B}_+(\mathbb{C}^d)}(F_V) = F_{V^\perp}$.
\end{ex}

\section{Main results}

From now on, we will consider two proper cones $K \subset L$ in real vector space $X$.

At the beginning, let us define a relation between an element $w \in L \setminus K \subset X$ and an element $\rho \in K^* \setminus L^* \subset X^*$.
\begin{df}[Detection of an element in $K^* \setminus L^*$ by an element in $L \setminus K$]
 We say, that an element $w \in L \setminus K$ detects an element $\rho \in K^* \setminus L^*$, iff $\rho(w)<0$. For an element $w \in L \setminus K$ we denote by $\mathcal{D}_{L|K}(w)$ the set of all states in $K^* \setminus L^*$ detected by $w$.
\label{detection}
\end{df}
The Banach separation theorem asures us, that for any element in $K^* \setminus L^*$ there exists an element in $L \setminus K$ detecting it, and that the dual fact holds, i.e. any element of the set $L \setminus K$ detects an element of a set $K^* \setminus L^*$. One can extend the definition \ref{detection} to the whole proper cone $L$ fixing $\mathcal{D}_{L|K}(k) = \emptyset$ for all $k \in K \subset L$.

For two elements $w_1, w_2 \in L \setminus K$ one can define a relation of ``being finer'':
\begin{df}[``Being finer``]
 We say, that an element $w_1 \in L \setminus K$ is finer than an element $w_2 \in L \setminus K$ with respect to the proper cone $K$, iff $\mathcal{D}_{L|K}(w_1) \supseteq \mathcal{D}_{L|K}(w_2)$ ($w_1$ detects more elements of $K^* \setminus L^*$ than $w_2$ in the sense of inclusion of sets). We denote this fact by $w_1 \ge_{f(K)} w_2$.
\label{beingf}
\end{df}
An element which is maximall with respect to this order is called optimal:
\begin{df}[Optimality]
 An element $w_1 \in L \setminus K$ is called \textit{optimal} with respect to the proper cone $K$, if there is no other element finer than $w_1$ in $L \setminus K$ (which detects more elements in $K^* \setminus L^*$).
\end{df}

On the other hand, one can define a relation of order with respect to the cone $K$:
\begin{df}
 An element $w_1 \in L$ is said to be greater than $w_2 \in L$ with respect to the cone $K$, iff
\begin{displaymath}
 \exists \lambda \in \mathbb{R}_+: \ w_1 - \lambda w_2 \in K.
\end{displaymath}
We will denote it by $w_1 \ge_K w_2$.
\end{df}

One can prove a theorem, that both following relations are equivalent. This is a generalization of Lemma 2 in \cite{opt} for arbitrary proper cones $L$ and $K \subset L$
\footnote{In \cite{opt} $L=\mathcal{W}_1$ and $K=\mathcal{B}_+$ or $K=\mathcal{S}_{PPT}$. Moreover, the work deals with normalized states and witnesses, what the reader should have in mind comparing results.}
.

\begin{twr}
 \begin{displaymath}
  w_1 \ge_{f(K)} w_2 \ \Leftrightarrow \ w_1 \le_K w_2
 \end{displaymath}
\end{twr}

\textbf{Proof:}
The proof bases on proof from \cite{opt}.

"$\Leftarrow$": Assume, that $w_1 \le_K w_2$. It means, that $w_1 = w_2 - k$ for an element of proper cone $K$. It means, that if only for an arbitrary $\rho \in K^*$ holds an inequality $\rho(w_2)<0$, then also $\rho(w_1) < 0$ holds, so $\mathcal{D}_{L|K}(w_1) \supset \mathcal{D}_{L|K}(w_2)$, and then $w_1 \ge_{L|K} w_2$.

"$\Rightarrow$": In the other side, assume that $w_1 \ge_{L|K} w_2$, so that $\mathcal{D}_{L|K}(w_1) \supset \mathcal{D}_{L|K}(w_2)$. We will prove, that $\lambda w_2 - w_1 \in K$, when a parameter $\lambda$ is chosen to be:
\begin{equation}
\lambda = \inf_{\rho \in \mathcal{D}_{L|K}(w_2)} \left| \frac{\rho(w_1)}{\rho(w_2)} \right|. 
\label{def_lambda_B}
\end{equation}
We will do it proving an inequality:
\begin{equation}
 \forall \rho \in K^* \quad \lambda \rho(w_2) \ge \rho(w_1)
\label{dow_opt_B}
\end{equation}
depending of the sign of the left-hand side.

\begin{enumerate}
 \item $\rho(w_2)=0 \ \Rightarrow \ \rho(w_1) \le 0$.

   Suppose, that for some $\rho \in K^*$ we have $\rho(w_2)=0 \ \land \ \rho(w_1) > 0$. Then there exists such an $\epsilon >0$, that $\forall \rho' \in B(\rho, \epsilon) \cap K^* \ \rho'(w_1) > 0$. This set contains unempty interior, so it have to consist states, for which $\rho'(w_2)<0$, but it denies the assumption $w_1 \ge_{L|K} w_2$.

 \item $\rho(w_2)<0 \ \Rightarrow \ \rho(w_1) \le \rho(w_2)$.

  We construct an element $\rho_1 \in K^*$ as $\rho_1 = \rho - \rho(w_2) I$, where $I$ denotes arbitrary element of a proper cone $K^*$, for which $I(w_2)=I(w_1)=1$. Such constructed $\rho_1$ fulfills the assumption from the previous case, so an equality $\rho_1(w_1) = \rho(w_1) - \rho(w_2) \le 0$ holds, what proves the postulated inequality.

  We will use it now to prove the inequality (\ref{dow_opt_B}) for $\rho(w_2)<0$. We know, that $\forall \rho \in \mathcal{D}_{L|K}(w_2)$ an inequality $\rho(w_1) < 0$ holds. It lets us to substitute the absolute value in the formula \ref{def_lambda_B} with negation, what leads to:
\begin{displaymath}
 \lambda = \inf_{\tilde{\rho} \in \mathcal{D}_{L|K}(w_2)} \frac{\tilde{\rho}(w_1)}{\tilde{\rho}(w_2)} \ \Rightarrow \ \frac{\rho(w_1)}{\rho(w_2)} \ge \lambda \ \Rightarrow \ \lambda \rho(w_2) \ge \rho(w_1),
\end{displaymath}
what proves the inequality \ref{dow_opt_B} in the case, when its left-hand side is negative.

 \item $\rho(w_2)>0 \ \Rightarrow \ \lambda \rho(w_2) \ge \rho(w_1)$

  Let's take an arbitrary element $\rho_1 \in \mathcal{D}_{L|K} (w_2)$. Let's define by use of it new element of the proper cone  $K^*$ as $\rho_2 = \rho(w_2) \rho_1 - \rho_1(w_2) \rho$. An equality $\rho_2(w_2)=0$ holds for it, so one can use to it the result of the first step an get $\rho_2(w_2)=\rho(w_2)\rho_1(w_1)-\rho_1(w_2) \rho(w_1) \le 0$. We get in result an inequality $\rho(w_2)\rho_1(w_1) \le \rho_1(w_2) \rho(w_1)$. Let's divide its sides by a negative number $\rho_1(w_2)\rho(w_2)$. We get then inequality:
\begin{displaymath}
 \frac{\rho_1(w_1)}{\rho_1(w_2)} \ge \frac{\rho(w_1)}{\rho(w_2)}
\end{displaymath}
The above inequality holds for an arbitrary $\rho_1 \in \mathcal{D}_{L|K}(w_2)$, so it holds also for the infimum of the right-hand side taken due to the set $\mathcal{D}(w_2)$. This infimum defines the $\lambda$. Multiplying both sides of such derived inequality by $\rho(w_1)$ one gets the inequality (\ref{dow_opt_B}) for $\rho(w_2)>0$.
\end{enumerate}
We have shown in this way, that the inequality (\ref{dow_opt_B}) is fulfilled independly on the sign of its left-hand side.
$\square$

\begin{ex}
 Choosing $\mathcal{W}_1$ as $L$ and $\mathcal{B}_+$ as $K$, one gets Theorem 1.
\end{ex}

\begin{ex}
 Choosing $\mathcal{W}_1$ as $L$ and $\mathcal{W}_D$ as $K$, one gets Theorem 2.
\end{ex}

Having proven the theorem, on has immediately the following:
\begin{pr}
 The element $w \in L$ is optimal with respect to $K$ iff $w-k \not \in L$ for any $k \in K$.
\end{pr}

\subsection{Geometrical properties of optimality}

\begin{lem}
 $F \subset \mathrm{opt}(L|K) \ \Leftrightarrow \ F \cap K = \{ 0 \}$
\label{geomchar}
\end{lem}

\textbf{Proof:} ''$\Leftarrow$'' Suppose, that $F \not \subset \mathrm{opt}(L|K)$. It means, that there exists an element $x \in F$ which is not optimal:
\begin{displaymath}
 \exists k \in K \ \exists l \in L: \ x = l + k \ \land \ k \ne 0.
\end{displaymath}
Face $F$ is an intersection of the cone $L$ and the kernel of some linear functional $\Phi_F$ which is non-negative on $L$. If $x \in F \iff \Phi_F(x)=0$, then $\Phi(l), \Phi(k)=0$. But it implies, that $l,k \in F_L(x)$.

''$\Rightarrow$'' A non-zero element of $K$ detects no elements in $K^* \setminus L^*$, so it cannot be optimal (any other set contains empty set - any other witness detects better). $F \subset \mathrm{opt}(L|K)$ only if all elements of $F$ are optimal, so if $F \cap K \ne \{ 0 \}$, $F \not \subset \mathrm{opt}(L|K)$.
$\square$

So all faces without elements from $K$ are included in $\mathrm{opt}(L|K)$. What could be said about the rest of faces - which contain an element from $K$?

\begin{lem}
 $F \cap K \ne \{ 0 \} \iff \mathrm{Int} F \cap \mathrm{opt}(L|K) = \{ 0 \}$.
\end{lem}
\textbf{Proof:} ''$\Leftarrow$''$w \in \mathrm{Int}(F)$ iff:
\begin{displaymath}
 \exists \epsilon > 0: \ B_{\ker \Phi_F}(w,\epsilon) \subset F
\end{displaymath}
(the topology on $F$ is the natural, metric topology of $\ker \Phi_F$ and the interior is understood due to this topology). It means, that:
\begin{displaymath}
 \forall y \in F \ \exists \epsilon>0: \ w + \epsilon(w-y) \in F.
\end{displaymath}
Now suppose that left-hand side holds, i.e. $F \cap K \ne \emptyset$. Let $k$ be an element of $F \cap K$. Then:
\begin{displaymath}
 (1+\epsilon)w - \epsilon k \in F
\end{displaymath}
so neither $(1+\epsilon)w$, neither $w$ is optimal.

''$\Rightarrow$'' If $w$ is optimal, and $w \in \mathrm{Int} F \iff \forall y \ \exists \epsilon>0: \ (1+\epsilon)w-\epsilon y \in F \subset L$, then no such an $y$ cannot be from $K$ (otherwise it would destroy the optimality of $w$).
$\square$

So if $F \cap K \ne \{ 0 \}$, then only optimal elements in $F$ can be in the boundary.

\vspace{0.5cm}

We will need the following helpful fact to prove the next lemma:
\begin{lem}
 Having $G \vartriangleleft K, F \vartriangleleft L$ for a pair of proper cones $K \subset L$:
 \begin{displaymath}
  G \subset F \iff \mathfrak{C}_K(G) \supset \mathfrak{C}_L(F)
 \end{displaymath}
 Observe also, that any $G$, which is an intersection of $K$ and a face $F \vartriangleleft L$ is a face of $K$
 \label{GF}
\end{lem}

\begin{lem}
 For an element $w \in L$ holds the following: $F_L(w) \subset \mathrm{opt}(L|K)$ if and only if $\mathfrak{C}_L(\{w\}) \cap \mathrm{Int}(K^*) \ne \emptyset$.
 \label{emptyint}
\end{lem}

\textbf{Proof:}
Fact, that $F_L(w) \subset \mathrm{opt}(L|K)$ ast equivalent (due to lemma \ref{emptyint}) to $F_L(w) \cap K = \{ 0 \}$. It can be written down in the following manner:
\begin{displaymath}
 \forall G \subset K \ \ G \subset F_L(w) \Rightarrow G = \{ 0 \}.
\end{displaymath}
and next using lemma \ref{GF}:
\begin{eqnarray*}
\forall G \subset K \ \mathfrak{C}_K(G) \supset \mathfrak{C}_L(F_L(w)) \Rightarrow G = \{ 0 \} \\
\forall H \vartriangleright K^* \ H \supset \mathfrak{C}_L(F_L(w)) =\mathfrak{C}_L(\{w\}) \Rightarrow H = K^*
\end{eqnarray*}
The only face in $K^*$ containing $\mathfrak{C}_L(\{w\})$ is the whole $K^*$, so $\mathfrak{C}_L(\{w\})$ generates the whole $K^*$ in $K^*$.  It is possible only if the set of generators contains the element from interior of $K^*$.
$\square$

The consequence of the above lemma is that the RHS implies the optimality of $w$. To get the opposite implication, one has to add an auxillary assumption, that $w \in \mathrm{Int}(F_L(w))$. For some cones, for example for cones with a polytope as a base set, $\forall w \in L \ F_L(w) = \mathrm{Int}(F_L(w))$ (the boundary of a face is a sum of faces) and then optimality of an element $w$ is equivalent to RHS of lemma \ref{emptyint}, but for general cones RHS does not imply optimality.



\begin{ex}
 Let $L = \mathcal{W}_1(\mathbb{C}^{d_1} \otimes \mathbb{C}^{d_2})$ and $K = \mathcal{B}_+(\mathbb{C}^{d_1} \otimes \mathbb{C}^{d_2})$. Then
 \begin{displaymath}
  \mathfrak{C}_L(w) = \{ \rho - \mathrm{separable}: \ \langle \rho| w \rangle = 0 \}
 \end{displaymath}
 Interior of $K^*$ is a subset of states for which $\det \rho = 0$. One can write down $\mathfrak{C}_L(w)$ as a convex hull:
 \begin{displaymath}
  \mathfrak{C}_L(w) = \mathrm{conv} \{ e \otimes f: \ \langle e \otimes f | w | e \otimes f \rangle \}
 \end{displaymath}
This convex hull contains a full rank state iff:
\begin{displaymath}
 \mathrm{span} \{ e \otimes f: \ \langle e \otimes f | w | e \otimes f \rangle \} = \mathbb{C}^{d_1} \otimes \mathbb{C}^{d_2}
\end{displaymath}
\end{ex}

This fact was presented in \cite{opt}, as a sufficient condition of optimality. It is known, that it is not neccesary condition for optimality and the witness $W_{Ch}$related to the Choi map in $\mathbb{C}^3 \times \mathbb{C}^3$ is a counter example - product vectors on which $W_{Ch}$ takes mean value zero spans only a 7-dimensional subspace of $\mathbb{C}^3 \times \mathbb{C}^3$. It tells us, that cones of entanglement witnesses do not have this wanted property, that boundary of a face is a sum of faces (which would let us make the sufficient condition the necessary condition).

Let's now consider the same condition for nd-optimality.

\begin{ex}
 Let $L = \mathcal{W}_1(\mathbb{C}^{d_1} \otimes \mathbb{C}^{d_2})$ and $K = \mathcal{W}_D(\mathbb{C}^{d_1} \otimes \mathbb{C}^{d_2})$. Then $\mathfrak{C}_L(w)$ is again spanned by the set of projectors onto product vectors $\phi \otimes \psi$, for which the inequality $\langle \phi \otimes \psi | w | \phi \otimes \psi \rangle = 0$ holds. The cone $\mathcal{S}_{PPT} = \mathcal{B}_+ \cap \mathcal{B}_+^\Gamma$, so the interior of $\mathcal{S}_{PPT}$ is the intersection of the interiors of $\mathcal{B}_+$ and $\mathcal{B}_+^\Gamma$. One has:
\begin{displaymath}
 \rho \in \mathrm{Int} \mathcal{S}_{PPT} \ \Leftrightarrow \ \rho \in \mathrm{Int} \mathcal{B}_+ \cap \mathrm{Int}
 \mathcal{B}_+^\Gamma \ \Leftrightarrow \rho \in \mathrm{Int} \mathcal{B}_+ \ \land \ \rho^\Gamma \in \mathrm{Int} \mathcal{B}_+
\end{displaymath}
The element $W$ is optimall iff it is possible to find in $\mathfrak{C}_L(w)$ a state $\rho$ of full rank, such that $\rho^\Gamma$ is also of full rank. It means that both following conditions are fulfilled:
\begin{eqnarray*}
 \mathrm{span} \{ e \otimes f: \ \langle e \otimes f | W | e \otimes f \rangle=0 \} = \mathbb{C}^{d_1} \otimes \mathbb{C}^{d_2} \\
 \mathrm{span} \{ e \otimes f: \ \langle e \otimes f^* | W | e \otimes f^* \rangle=0 \} \\
 = \mathrm{span} \{e \otimes f: \ \langle e \otimes f | W^\Gamma | e \otimes f \rangle = 0 \} = \mathbb{C}^{d_1} \otimes \mathbb{C}^{d_2}
\end{eqnarray*}
What means that $W$ is nd-optimal iff both $W$ and $W^\Gamma$ is optimal.
\end{ex}


\ack

The whole theory was born in very fruitfull discussion with Jarek Korbicz and Darek Chru\'sci\'nski.

This work was partially supported by the
Polish Ministry of Science and Higher Education Grant No
3004/B/H03/2007/33.

\end{document}